\documentclass{IEEEcsmag}

\usepackage[colorlinks,urlcolor=blue,linkcolor=blue,citecolor=blue]{hyperref}

\usepackage{upmath}

\jvol{XX}
\jnum{XX}
\paper{8}
\jmonth{May/June}
\jname{IT Professional}
\pubyear{2019}

\usepackage[numbers]{natbib}
\usepackage{graphicx}
\usepackage{caption}
\usepackage{subcaption}
\usepackage[export]{adjustbox}
\usepackage{xcolor}
\usepackage{soul}
\usepackage{multirow}
\usepackage[final]{changes}
\usepackage{comment}
\definechangesauthor[name={add}, color=red]{add}

\begin{document}

%\sptitle{Department: Head}
%\editor{Editor: Name, xxxx@email}

\title{Towards personalised music-therapy; a neurocomputational modelling perspective}

\author{Nicole Lai}
\affil{University of Glasgow}

\author{Marios Philiastides}
\affil{University of Glasgow}

\author{Fahim Kawsar}
\affil{University of Glasgow}

\author{Fani Deligianni}
\affil{University of Glasgow}

%\markboth{Department Head}{Paper title}

\begin{abstract}
Music therapy has emerged recently as a successful intervention that improves patient's outcome in a large range of neurological and mood disorders without adverse effects. Brain networks are entrained to music in ways that can be explained both via top-down and bottom-up processes. In particular, the direct interaction of auditory with the motor \added[id=add]{and the reward system} via a predictive framework explains the efficacy of music-based interventions in motor rehabilitation. In this manuscript, we provide a brief overview of current theories of music perception and processing. Subsequently, we summarise evidence of music-based interventions primarily in motor, emotional and cardiovascular regulation. We highlight opportunities to improve quality of life and reduce stress beyond the clinic environment and in healthy individuals. This relatively unexplored area requires an understanding of how we can personalise and automate music selection processes to fit individuals needs and tasks via feedback loops mediated by measurements of neuro-physiological responses.  

\end{abstract}

\maketitle

\chapterinitial{Music is universal across cultures} and similar to language it has expressive and motivational roles. Fundamental properties in music composition is the presence of a rhythmic structure and in music perception is \added[id=add]{one's} innate ability to synchronise to the rhythm, called \textbf{sensorimotor synchronisation (SMS)} \cite{Tichko2022a}. Rhythm and synchronisation also facilitate human communication by modulating attention, which is observed even in infant's perception and response to musical stimulus \cite{Tichko2022a}. It is intriguing that several neuro-developmental and neurological disorders are linked to deficiencies related to rhythm, timing, and synchrony processing \cite{lense}. In fact, rhythm and time perception involves interactions between the auditory and motor systems.

The use of music has become an integrated tool in the daily lifestyle for adults via streaming platforms and portable devices that can provide personalised playlists according to \added[id=add]{user's} music preferences. Hence \added[id=add]{it's} therapeutic benefits and application in a clinical setting has been \deleted{recently} confirmed in well-controlled randomised trials \cite{ThautC}. In order to optimise and personalise the efficacy of these methods to specific patient groups and to improve well-being in healthy adults, it is imperative to understand which specific characteristics in music can give rise to its therapeutic benefits. Equally important is to enable biofeedback via intelligent algorithms that take into account the neurophysiological responses of the individual. 

The application of music to support some form of rehabilitation in a clinical setting, known as music therapy, was considered to be a social science. However, with recent advances in cognitive brain imaging, models of music perception have been developed that can benefit both patients and healthy humans. Understanding the effects of music on brain dynamics can help us shape music therapy techniques.  In this review, we summarise the key concepts behind \deleted{computational} models \added[id=add]{of musical perception and processing} that underpin current evidence from brain imaging studies and explain the efficacy of music-based therapy in a broad range of clinical and psychiatric disorders.  \added[id=add]{Firstly, we discuss how} communication between the auditory-motor networks and external rhythmic cues in an \textbf{oscillatory neurodynamics model} achieves a sufficient level of entrainment as demonstrated by synchronisation to the beat of the music \cite{Large2015}. \added[id=add]{We highlight Neural Resonance Theory (NRT) as a generalised framework of describing music perception as dynamic systems of coupled neural oscillators. This formulation allows the incorporation of developmental principles by changing the strength of coupling between oscillators to reflect Hebbian Plasticity. Finally, we explore three major formulations of predictive-coding models, namely the Predictive Coding Model of Music (PCM), the Predictive Coding model of Rhythmic Incongruity (PCRI) and the Action Simulation for Auditory Prediction (ASAP). 
In these models, musical perception is the result of expectations, generated from prior experiences and knowledge, compared to the sensory input.} \deleted{A combination of a predictive and oscillatory models is proposed in the \textbf{dynamics of attending} model. The model suggests that the listener's perception is continuously updated as a result of an ability to live track temporal fluctuations} \deleted{Tichko et al's., 2022 \cite{Tichko2022a} proposed developmental theory grounded in Neural Resonance Theory (NRT) and the addition of Hebbian Plasticity formed a non-representation or ``radically-embodied'' approach to music processing.} 

Subsequently, we summarise current evidence of music-based interventions in neurorehabilitation, as well as regulation of affective states and cardiovascular function.  
Finally, we highlight key challenges in automating and personalising music-based interventions in the community. A vital consideration of the implementation of music therapy is that positive effects are limited in duration. In other words, in patients with movement disorders, for example, the improvements in gait parameters will decrease if the treatment is discontinued. Current evidence suggests that music therapy could be an ideal intervention to apply in the community with patients being able to continue practicing exercises with minimal equipment and effort to execute. Nevertheless, it also highlights the importance of ensuring that the intervention is supported such that patients are encouraged to continue to maintain the beneficial rewards. 

\section{Models of Music Perception and Processing}

Early works on understanding how music affects the brain and its potentially therapeutic benefits have focused on rhythm and entrainment. Entrainment is a universal phenomenon that describes the process of interaction between independent rhythmical processes. \deleted{In the event of an external perturbation, when two rhythms are entrained, it is expected that both would maintain a stable phase relationship whilst returning to their original relationship.} One example is in the coupling the musical beat and index tapping as modelled according the Haken-Kelso-Bunz (HKB) model \cite{Roman}. Importantly, entrainment does not necessarily require in-phase synchronisation between rhythms of matching periods. Different rhythms can entrain in hierarchical or polyrhythmical relationships. In particular, the levels of the oscillatory hierarchy can be reflective of the rhythmical structure, such that the period of the oscillation is proportional to each level of note duration. Polyrhythm is the simultaneous presence of two or more rhythmically contrasting rhythms in which an unequal number of beats is spaced out equally within the same measure, for example three beats, known as triplets, against two beats (three-against-two). This can be represented in the oscillatory model as lower levels of the rhythmic structure may exist in anti-phase with one another however be in-phase with the top-level creating the polyrhythmical structure. 

\deleted{Thaut et al. was the first to demonstrate neural entrainment in neurorehabilitation, showing that auditory rhythmic patterns could entrain movement patterns in patients with movement disorders} \added[id=add]{Neural entrainment has been shown to be evident in rehabilitative techniques, for example auditory rhythmic patterns can entrain movement in patients with motor disorders \cite{Janzen}}. Although, the biological basis of neural entrainment is not fully understood, temporal modulation of the beta band (13-35 Hz), which dominates oscillations in sensorimotor and motor areas of the brain is thought to play a key role \cite{Janzen}. The phase relationship is maintained even after the stimulus stops, which provides an anticipatory mechanism that is used to predict the next beat. 

These observations have led to an \textbf{oscillatory neurodynamics model} representation of nonlinear brain responses induced by musical stimulus \cite{Large2015}. The oscillatory neurodynamics model assumes that rhythm is a hierarchical structure of coupled `active, self-sustained oscillator' \cite{LargeP}. The oscillator's role is to synchronise with the external rhythms such that the phase of the oscillator becomes a representation of the beat onset in the rhythmic sequence. The coupling of multiple oscillators gyrating at different periods, represents the different temporal levels of the metrical structure, with the time at which the coinciding peaks of the different levels of oscillators represents the ‘strong’ beats. Furthermore, there exists a `sensitive region' within the oscillator's period acting as a representation of the likelihood of the beat falling within that time period. As a result of this region, temporal fluctuations that occur in natural performances are processed without it affecting the perception of the overall tempo. Large and Palmer \cite{LargeP} indicate that it is the coupling of the multiple oscillators that provides the support in its beat-tracking ability. It represents the relationship between metrical levels such that one can stabilise another should it begin deviating \cite{LargeP}. In other words, the model provides a computational depiction of our ability to overlook temporal fluctuations for expressive quality, whilst also adapting our beat perception if the beat is consistently heard outside of our ‘sensitive region’ of prediction. 

\added[id=add]{Closely linked with the oscillatory neurodynamic model, \textbf{ the dynamics of attending (DAT) theory} specifies entrainment between attending rhythms to the auditory rhythms to track time-varying events \cite{dynamics_of_attending}. In particular, attentional focus increases at points that are predictable and temporally regular \cite{dynamics_of_attending}, therefore enhancing processing. Subsequently, the neurodynamical model is perceived as a theory of prediction. Predictions are `actively made' in an anticipatory manner as a result of it's time-delay characteristics \cite{Roman}, for example listeners' are able to anticipate the next beat before it is heard \cite{Janzen}}.

\added[id=add]{The \textbf{Neural Resonance Theory (NRT)} defines entrainment as coupled neural oscillators that obey the laws of mathematical dynamical systems of oscillation such that they are \textit{physical} oscillations and not a representation of the external musical rhythm \cite{Tichko2022a}. Tichko et al., 2022a \cite{Tichko2022a} proposes a developmental theory grounded in NRT with \textbf{Hebbian Plasticity} to present a dynamical system of perception that encompasses neurodevelopmental principles.  The external musical oscillations is proposed to resonate with the bio-physical oscillations in a law-governed manner, such that structural regularities become established in perception, action and attention \cite{Tichko2022a}. The additional process of Hebbian Plasticity serves to attune the oscillations with the environment through adjustments of the coupled oscillators. Development of dynamic memories of complex rhythmic patterns are the result of changes in amplitude and phase relationship between coupled oscillators, representing the strength of the synaptic connection of `neurons that wire together, fire together'. Application of this theory to model infant's perception and development of musical rhythm have proven that the oscillatory neurodynamics model encompasses both bottom up (perception is built solely on received sensory information from environment) and top-down processes (prior knowledge and experience is used to shape perception and interpret sensory inputs). Infants that received auditory-vestibular training, in the form of bouncing, shaped their neural response when tested on unaccented rhythms, suggesting that a combination of multiple sensory inputs will elicit stronger resonance. Developmental plasticity was reflected in stronger auditory-vestibular oscillatory connections achieved during the training phase \cite{Tichko2022a}.}

 \deleted{a bottom-up approach that explains fundamental principles of the computational neuroscience behind neuronal entrainment via generative models of brain oscillatory activity. In contrast, the Predictive-coding of musical characteristics such as melody, rhythm and harmony and it's link to music perception, action, learning and emotions illustrates music listening from a top-down perspective. As expected from it's name, the predictive-coding model is one of prediction theory, however the nature of predictability differs from that of the neurodynamical model previously discussed.} \deleted{top-down approaches of linking music perception, action, learning and emotions, predictive coding of musical characteristics such as melody, rhythm and harmony.}
 Whilst, the oscillatory neurodynamics model \deleted{is considered} \added[id=add]{explains musical perception under mathematical dynamic systems of neuronal entrainment}, the \textbf{Predictive coding of music model (PCM)} \deleted{extends the PCRI model beyond rhythm to encompass melody and harmony} \deleted{It} exploits active inference, stating that perception, action and learning can be modelled as a recursive Bayesian process. The predictive-coding of musical characteristics such as melody, rhythm and harmony and it's link to music perception, action, learning and emotions illustrates music listening from a top-down perspective. Expectations are generated from previous experience and  \deleted{also} \added[id=add]{therefore} \textit{inferred} by contextual factors such as cultural background, musical competence and individual traits \cite{Vuust}. \deleted{Predictions of rhythmic properties, for example beat detection, are then made based on minimising the prediction errors between expected and actual rhythmic perception.} \textbf{Prediction errors}, defined as the difference between the \textit{predicted} and the \textit{actual} sensory input, are attempted to be minimised through the statistical weighting of each prediction error based on their expected \textit{precision} \cite{Vuust}. The internal model is proposed to be constructed of a \textit{hierarchical structure} of networks such that prediction errors ascend through the levels in order to update higher levels of processing and result in predictions being cascaded down to resolve lower-level prediction errors \cite{Koelsch2019}. This top-down and bottom-up communication serves to continuously update the model and therefore minimises variational \textit{free-energy} \cite{Cheung2019}. \textbf{Event-related potentials (ERPs)} such as the \textbf{mismatched negativity (MMN)}, particularly it's amplitude can be used as preattentive markers, of predictive errors \cite{Vuust}. MMN occurs as a response to an auditory ``oddball'' or deviant from the established structure.

The \deleted{ASAP hypothesis led to the} \textbf{Predictive Coding of Rhythmic Incongruity (PCRI) model} \cite{Vuust}, \deleted{which models the brain behavioural response to rhythm as a weighted prediction error.} \added[id=add]{is introduced as a formal application of the PCM for musical rhythm exclusively}. The PCRI model explains the feelings of ``wanting to move with the music'', known as groove, under the experience of `incongruent' note onset that disturb the regular flow of rhythm (syncopation) and the listener urge to `correct' these prediction errors \cite{Vuust}. Neuroimaging findings support the pleasurable notion of `groove' with activations in the motor and reward networks including the basal ganglia \cite{Matthews2020}.

The \textbf{Action Simulation for Auditory Prediction (ASAP)} hypothesis suggests that \deleted{prediction plays a key part of rhythm perception} `pure' beat perception also involves generations of simulated motor actions \cite{ASAP}, despite no physical movement.  The ASAP model \added[id=add]{explains} \deleted{is built on the argument that musical beat is predictable by the listener, emphasising a top-down process,} the listener's readiness to ‘move’ \cite{Janzen} as a result of a predictive process. Patel and Iversen \cite{ASAP} hypothesise that there exists a two-way relationship between the auditory and motor planning regions. Specifically, the temporal information of auditory stimuli is initially received by the auditory regions and communicated to the motor region, whereby the timing of this communication influences the period of the signals within the motor planning regions. This in turn allows neural activity within the motor cortex to be entrained to the beat, known as the `simulated action'. Finally the motor regions return this signal to the auditory regions to facilitate predictions of subsequent incoming beat times \cite{ASAP}. In other words, the motor planning region of the brain facilitates the perception of musical beats by `simulating' the periodic `action'.

\added[id=add]{One limitation of the theory is that it} does not explain how our perception is translated into physical movements or the computations that take place behind the predictions \cite{Large2015}. Instead it is assumed that the movements are a natural consequences of the motor planning regions influencing nearby regions responsible for the governing of movement \cite{ASAP}. \deleted{Additionally, whilst the model provides a detailed approach to the cognitive structures involved in overall rhythm perception and its complementary effects on emotional stimulation, it does not provide an explanation for the computations that take place behind the predictions} In other words, the model does not consider our ability to detect the beat amongst the rhythm or give weight to the listener’s ability to isolate the pulse within a rhythmic sequence, unlike in the neurodynamical models previously described. Both the predictive-coding approach and the neurodynamical model individually do not provide a comprehensive account of musical perception from a neurological perspective. Vuust et al. \cite{Vuust} proposes that a combination of the two models could provide a more satisfactory modelling of our musical perception and processing. 

\deleted{The \textbf{Dynamics of Attending Theory} proposes such a combination which explains how listener's perception adapts to events that changes over time. It builds on both the notion of entrainment and expectancy to describe tracking of time-varying events in real-time. The internal oscillations, termed as \textbf{attending rhythms} occurs between the auditory-motor systems and generates expectancies therefore creating the predictions for future events as in the ASAP model. Nevertheless the attending rhythms are driven by the external rhythm to achieve entrainment, with the addition of a \textbf{stable attractor} that allows adjustments to momentary changes. The attractor is responsible for the long-term control over the attending rhythm such that if an unexpected onset occurs in the external rhythm, stability of the perturbed attending rhythm is reached similar to the role of the couple oscillators.
A graphical representation of our rhythm perception as a combination of the predictive and oscillatory model described can be seen in Figure \ref{model}.}

Another intriguing observation is that predictable music is not necessarily associated with pleasure \cite{levitin} - instead this is most commonly met with feelings of annoyance. \added[id=add]{One resolution to this conundrum rests upon music's `epistemic offering' \cite{Koelsch2019}.  Cheung et al, 2019 \cite{Cheung2019} proposes that the retrospective and prospective states of expectation - surprise and uncertainty respectively, are important additions in understanding the role of musical pleasure in the predictive-coding model.  Activations of the nucleus accumbens (NAcc) and the use of the dopaminergic pathway was proposed to facilitate musical pleasure by directing the listener to `want to find out what happens next' in order to resolve the uncertainty in the music. Cheung et al., 2019 \cite{Cheung2019} proposes that the level of dopamine release represents the precision or inverse of uncertainty, of the prediction errors in the free-energy principle.} 

The threshold of pleasurable `surprise and uncertainty' can be modelled by a power law of temporal fluctuations, as proposed in the \textbf{1/\textit{f} distribution} \cite{levitin}. The conclusion deduced from a large collection of compositions ranging across four centuries and several genres of Western music that obeyed this 1/\textit{f} rhythmic structure suggests that this fractal-relationship could be key to accomplishing the `right level' of synchronisation for enjoyment. \added[id=add]{Despite, the `uncertainty' that is proposed to derive pleasure in fractal structure, listeners are still able to anticipate temporal fluctuations as modelled by the neurodynamical model \cite{Rankin2014}. This anticipatory behaviour is suggested to be guided by the long-term structures and not dependent on musical skill or training \cite{Rankin2014}}.

\section{Music in Neuro-rehabilitation and regulation of emotional and cardiovascular systems}

One of the most natural and inherent response to music is to synchronise our bodily movements to the rhythmic structure. The effortless and often subconscious onset of movement is an example of the deeply ingrained relationship between the motor and auditory processing networks \cite{ASAP}. Therefore, it is not surprising that this biological relationship has been exploited in the field of motor rehabilitation \cite{Moumdijian2018}. In neurological disorders, despite continuous pharmacological advances, gait/motor impairments notably can be resistant to medication, often only contributing to the symptomatic relief as opposed to remedying the underlying pathology \cite{Ghai}. Furthermore, with all pharmacological interventions, there can be harmful side-effect to health and quality of life, for example  toxicity and loss of drug efficacy \cite{Ghai}. 

One branch of music therapy that centres around neurorehabilitation is \textbf{Neurologic Music Therapy (NMT)} and is defined as `the therapeutic application of music to cognitive, sensory and motor dysfunction due to neurological disease of the human nervous system' \cite{ThautC}. An example of a NMT used for cognitive rehabilitation is \textbf{Music Sonification Therapy} that has been frequently employed in stroke rehabilitation. Sonification is broadly defined as the use of non-speech audio to represent information \cite{Scholz}. The musical intervention improves on motor abilities that have been affected by a neurological disorder such as stroke by retraining gross motor functions through the use of repeated movements of the affected limb \cite{Scholz}.

One of the most common technique to obtain gait harmony in music therapy is the application of \textbf{Rhythmic Auditory Stimulation (RAS)}. RAS takes advantage of our natural driving force to move as a response to listening to music and utilizes the coupling of the auditory-motor networks with the aim to promote a stable and adaptive walking pattern in patients with gait deficits as a result of some neurological condition such as stroke, Parkinson's disease or traumatic brain injury or general effects of ageing \cite{Sihvonen}. Characteristics of an abnormal gait patterns include shuffling, decrease in walking speed, shorted stride length and asymmetric stride durations \cite{Ghai} which often leads to high risk of falls. RAS is built on entrainment of motor functions to the external timekeeper in the musical stimuli. Synchronisation between the external beat and steps indicates a phase-lock and entrainment of the auditory and motor system \cite{Moumdijian2018}.

\begin{figure*}[!tbp]
 \begin{subfigure}[b]{0.5\textwidth}
    \includegraphics[width=\textwidth, height=5cm]{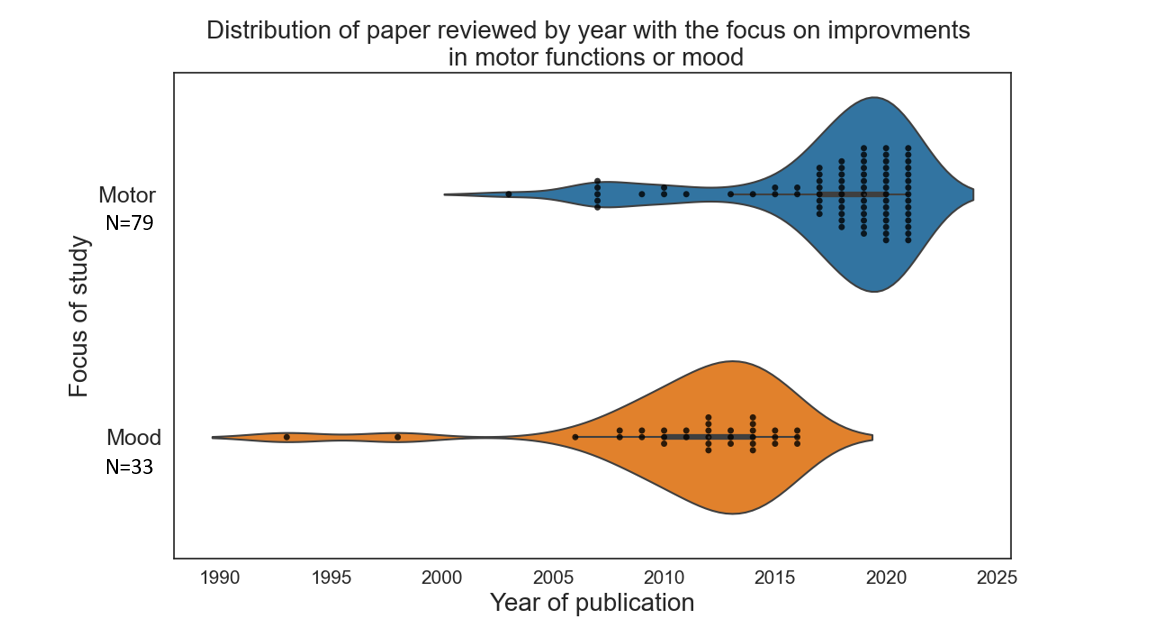}
    \caption{Distribution of publication years and primary focus of the study.}
    \label{violinplot}
  \end{subfigure}
  \hfill
  \begin{subfigure}[b]{0.5\textwidth}
    \includegraphics[width=\textwidth]{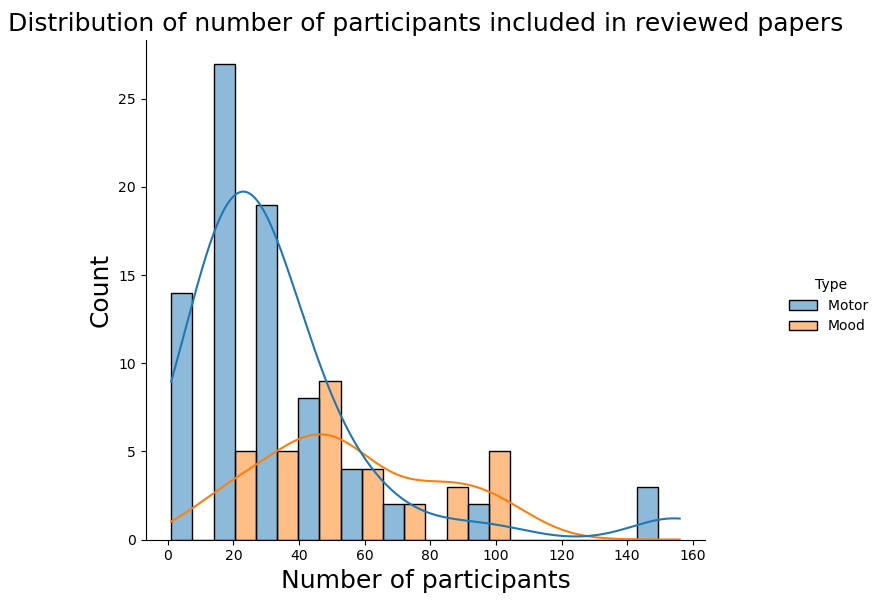}
    \caption{Distribution plot of the total number of participants.}
    \label{displot}
  \end{subfigure}
  \caption{Distribution of the publication year and number of participants in papers reviewed in \cite{Janzen}, \cite{Sihvonen}, \cite{Steen}}
  \label{stats}
\end{figure*}

\begin{figure*}[!tbp]
 \begin{subfigure}[b]{\textwidth}
    \includegraphics[width=\textwidth]{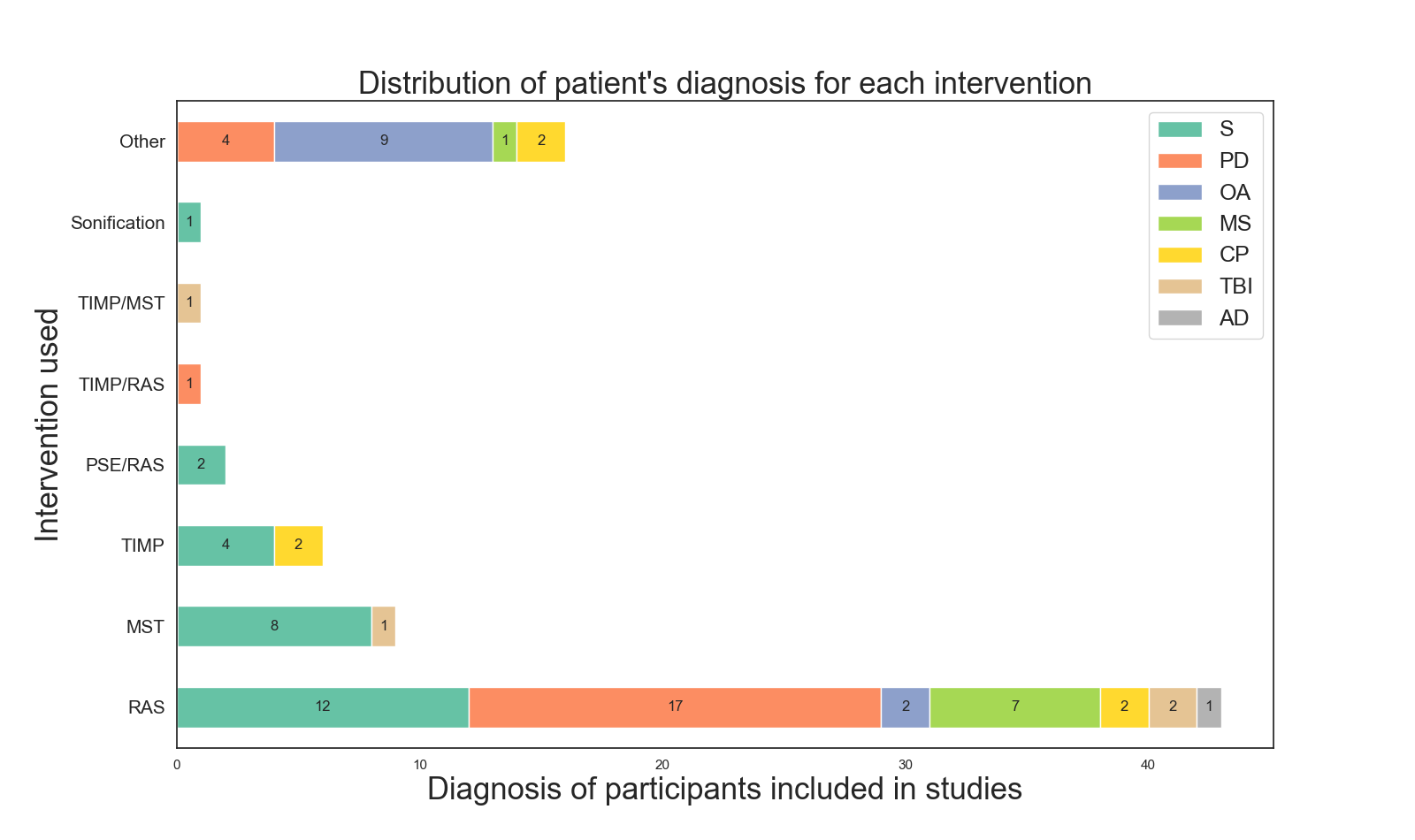}
    \caption{Discrete distribution of participant's diagnosis against interventions used in \cite{Janzen}, \cite{Sihvonen}
    \\
    RAS = Rhythmic Auditory Stimulation, MST = Music-supported Therapy, TIMP = Therapeutic Instrumental Music Performance PSE/RAS = Patterned Sensory Enhancement/Rhythmic Auditory Stimulation, RAS/TIMP = Rhythmic Auditory Stimulation/Therapeutic Instrumental Music Performance, TIMP/MST = Therapeutic Instrumental Music Performance/Music-supported Therapy. `Other' included studies that either did not specify the type of music intervention, instrumental training or playing and/or music listening.
    \\
    Further description on each intervention can be found in \cite{ThautC} and \cite{Janzen}}
    \label{diag_intervention}
  \end{subfigure}
  \hfill
  \begin{subfigure}[b]{\textwidth}
     \centering
    \includegraphics[width=\linewidth]{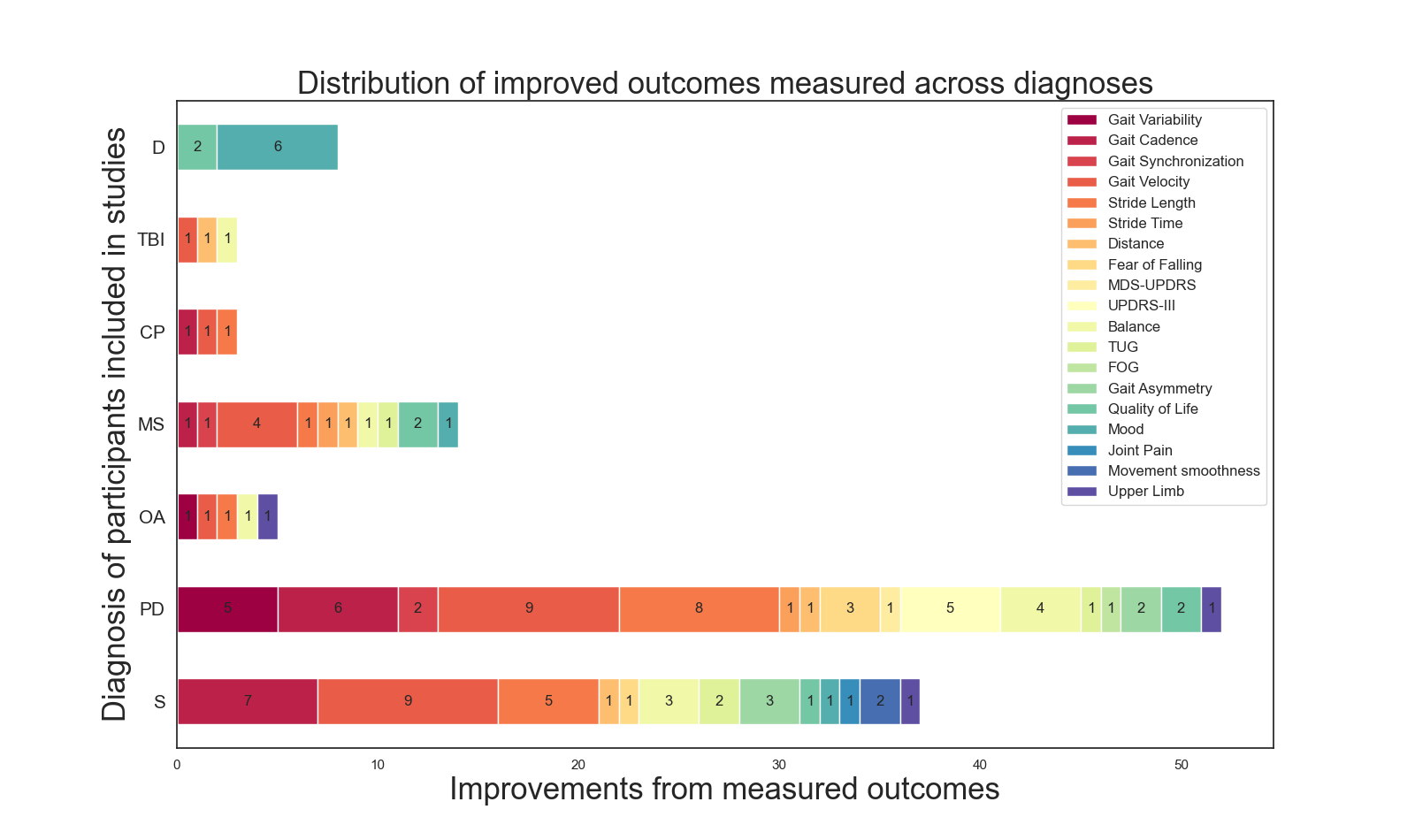}
    \caption{Discrete distribution of improved outcomes against participant's diagnosis  in \cite{Janzen}, \cite{Sihvonen}, \cite{Steen}. 
    \\
    S = Stroke, PD = Parkinson's Disease, OA = Older Adults, MS = Multiple Sclerosis, CP = Cerebral Palsy, TBI = Traumatic Brain Injury, D = Dementia.}
    \label{diag_measure}
    \end{subfigure}
    \caption{Discrete distributions of papers reviewed in \cite{Janzen}, \cite{Sihvonen}, \cite{Steen}}
\end{figure*}

There is a large body of research underpinning the use of rhythm and music-based interventions for neuro-rehabilitation of a large range of neurological and neuro-developmental disorders. Recent randomised controlled trials have shown that music-based interventions are effective in motor rehabilitation in a variety of conditions \cite{Janzen}. Seebacher et al.  \cite{Seebacher2016} reports that RAS during walking results in significant improvement\added[id=add]{s} in cognitive and physical fatigue as well as quality of life of patients with multiple sclerosis. Tong et al., \cite{Tong} showed significant improvement in post-stroke patients upper-limb motor function. Sihvonen et al. \cite{Sihvonen} also summarises results from 16 randomised trials in stroke-related neurological disturbances, which showed significant improvement in gait training with musical support. Music therapy has also resulted in improvements in speech and cognitive recovery for stroke patients \cite{Sihvonen}. In patients with Parkinson disease, music therapy with a coupling between movement and music has shown consistent and clinically significant results in improving motor symptoms \cite{Sihvonen}. Exposure to music has also showed a significant reduction in seizures in epileptic patients \cite{Sihvonen}. For detailed reviews on the music-based interventions the reader can refer to Janzen et al. \cite{Janzen}, which provides a comprehensive summary of studies on music and motor rehabilitation, whereas Sihvonen et al. \cite{Sihvonen} provides a review on music interventions in neurological rehabilitation and Steen et al. \cite{Steen} summarises current progress in music interventions for dementia. A graphical summary of the distribution of publication years and number of participants included in each study can be found in Figure \ref{stats}. Furthermore, Figure \ref{diag_intervention} and \ref{diag_measure} illustrates the distribution of firstly, the intervention used and the participants diagnosis and secondly, the diagnosis against the improvements observed. 

Music also has a strong established anthropological and sociological role in society, conventionally associated mood-regulating properties \cite{ThautC}. It is commonly accepted as one of the strongest tools used to arouse an affective state or provoke emotional response \cite{Chaturvedi}, with this property being one of the most cited reason to why listeners value music \cite{Lonsdale}. 
There are strong evidence that suggest that music-interventions can improve the outcome of patients with depression as well as alleviate depressive symptoms interleaved with other neurological disorders such as dementia and Parkinson. A recent meta-analysis of seventeen randomised control trials with patients with dementia suggests that music-based interventions improved symptoms of depression whereas it was unclear whether it improved the overall quality of life, agitation and cognition \cite{Steen}.

One goal of music listening, in a therapeutic context, is to reduce stress and increase feelings of relaxation that can be achieved via the engagement of the parasympathetic branch of the autonomic nervous system \cite{Leggieri2019}. Kulinski et al. \cite{Kulinski2022} provides a summary of several studies that show how cardio-respiratory variables, such as heart rate (HR), heart rate variability (HRV), blood pressure (BP) and respiration are modified with music-based interventions. Evidence also show that music has positive effects in physical performance and endurance.  

Music-based interventions can also have a profound effect in work environments. Occupational stress and mental health problems is considered as a ‘Health Epidemic of the 21st Century’ from WHO organisation \cite{Axelsen2022}. Large experiments on more than 600 participants in occupational settings demonstrated that just 10 minutes of music listening improved working memory along with self-perceived stress and measures of sustained attention \cite{Axelsen2022}. 

\section{Shaping Music-based Interventions}

Music emerges as a powerful way to improve health outcomes and well-being without pharmacological interventions in both patients and healthy people. 
One of the key principle\added[id=add]{s} of music therapy is that it is tailored to the \added[id=add]{participant's} condition, goal, abilities \cite{Janzen} such that it is not a `one-size-fits-all approach'. Therefore, one challenge is selecting the musical features that would optimise the efficacy of the music therapy. For example, the initial cadence or tempo is an important factor in the effects of the intervention in gait rehabilitation. Too slow of an internal cadence can lead to prolong time for a movement to be made and increases the danger of motion reinvestment, which implicates conscious control \cite{Ghai}. Automatic motor processes become disrupted and additional attentional resources are then required \cite{Ghai}. On the other hand, too fast of an initial tempo can lead to exceeding the participant's current physical capabilities which could lead to worsening conditions, feeling of disheartening and demotivation and place the participants into a state of high-stress \cite{Ghai}. In RAS, the rhythmic cues are chosen to coincide with the participant's initial cadence and incrementally increased or decreased by 5-10\% once the movement speed is entrained \cite{Janzen}. By setting the baseline as the participant's current or natural frequency, this enhances kinematic stability promoted by rhythmic entrainment as the initial movement parameters are optimised and stabilised \cite{ThautC}. Furthermore, in a review conducted by Janzen et al. \cite{Janzen} it was commented that the inconsistency of the effect of auditory cue on stride length variability across different studies could be due to the result of not tailoring initial cadences to the participant's baseline. They highlight the demand for additional research into optimising RAS cadence individually and taking into consideration external factors such participant's previous musical training and rhythmic abilities \cite{Janzen}. 

In the most fundamental form, musical \added[id=add]{stimuli} are often simply metronome beats which yield promising results in improving spatiotemporal gait parameters, such as increase stride length, velocity, cadence and symmetry \cite{Moumdijian2018}. However, these positive effects not permanent as shown by a significant decrease in spatiotemporal gait parameters (velocity and stride length significantly decreasing) once the RAS-training was discontinued \cite{Thaut2019}. This emphasises the need for a continuous intervention so that it can be implemented in a community setting for patients to continue training in their own homes for these effects to be maintained. Janzen et al. \cite{Janzen} conducted a withdrawal/discontinuation design study in which when the participants entered an 8-week discontinuation of RAS-training, during which the number of falls significantly increased and significantly decreased again once RAS-training resumed in week 16. 

Incorporating an emotional-motivational quality of music as a desirable secondary effect could be the solution in enhancing music therapy as a long-term intervention \cite{ThautC}. Key features that have been explored are level of beat perception, familiarity and level of groove perceived, often characterised by strong, salient beats \cite{familiarity}\cite{groove}. Higher levels of beat salience reduces the need for beat finding therefore reducing cognitive demands of synchronisation during RAS and positive effects on stride velocity and length. 

\added[id=add]{Familiarity modulates musical pleasure by exploiting the anticipation built from cultural or personal influence from previous exposure. In other words, a sense of predictability or familiarity is perceived to be pleasurable \cite{Salimpoor2013}. Musical pleasure derived from familiar music is correlated with increased blood oxygen level dependence (BOLD) in the reward systems, including the NAcc and putamen \cite{Pereira2011}. It can be argued that the mechanisms that underlie the effectiveness of music-based interventions in episodic memory are driven by the reward brain network \cite{Ferreri2022}.} \added[id=add]{Familiarity in the musical pieces can trigger long-term autobiographical memories unique to each listener and revive feelings associated to the people, location and conditions that were connected to the experience as shown in activations in the hippocampus \cite{Leggieri2019}. This factor can play an important role in neuro-degenerative processes that affect episodic memory such as in dementia and Alzheimer. Tichko et al., 2022b \cite{Tichko2022b} observed strong neural entrainment in older adults listening to self-selected music, suggesting that familiarity and the associate pleasure in music listening can aid in preserving neural connections employed in rhythmic perception that may have deteriorated as a result of aging. Additionally, there are strong evidence that vocal music listening enhanced verbal memory recovery in stroke patients that suffer with aphasia \cite{SihvonenVocal}, in addition to improvement in anxiety and depression \cite{Leggieri2019}.}

One prevailing challenge when exploring the benefits of listening to music, in particular the listener’s emotional response, is that often researchers are reliant on the use of subjects' feedback through self-reports that maps the stimulus to a small set of predefined emotions \cite{deligianni2019}. These are subjective and influenced by social and cultural norms, often only taking into consideration four to five types of discrete emotions, which is difficult to obtain regularly with relation to the music time scale. Subsequently as the result of the reliance on subject's feedback through self-reports when listening to music, there is a difficulty in distinguishing and validating if these reports are of felt or perceived emotions. 

Other methods include measuring physiological data including electrodermal activity, blood volume pulse, skin temporal and pupil dilation. For example decreases in heart and respiratory rates can signal a reduction in feelings of stress and anxiety \cite{Rahman}. Although, physiological data can provide a measure of emotional arousal, they are arguably an indirect measure of the autonomous nervous system function, which might undermine the effects of music entrainment to other major brain networks. Perhaps, a combination of physiological recordings with cognitive screening might be able to result in more accurate models of brain function modulation via brain computer interfaces. In this way, research can shed light on which elements of music structure have a significant effect in music-based interventions.

There exists a compromise between the ability to accurately record brain activity and carry out experiments such that they are ecologically valid. The integrated use of music in daily lifestyle suggests we do not use music solely as an entertainment purpose but also as a secondary function as aiding concentration on the main task, for example whilst studying or working. Students report that music enhances concentration and it also encourages motivation and enjoyment of completion of the main task \cite{Lonsdale}. In the same way that the choice of music is important for effective neurorehabilitation, the choice of background music is critical to prevent distraction and undesirable effects. Again, the deciding factor in what makes a music 'good for concentration' seems to point to the listener's preference and an optimal level of enjoyment in comparison to other aspects such as genre \cite{Yakura}. 
This has led to the proposition of automated recommending systems of background music while working. For example, FocusMusicRecommender \cite{Yakura} creates an automatic playback function to support in maintaining concentration on the main task by taking away the need for the user to select songs during work. 

There is a compromise between modulate levels of enjoyment that would motivate the listener to focus on the main task and high levels of arousal that could distract them. Therefore it was recommended to focus on selecting songs that user's may not like or dislike in an aim to maintain concentration \cite{Yakura}. 
Only recently it was recognised the need to personalise therapeutic music to an individual needs \cite{fahim2022}. It remains elusive how this can be measured outside laboratory settings. One suggestion is the use of crowdsourcing to evaluate the use of music interventions in the workplace \cite{Axelsen2022}. This involves mobile apps that measure sustained attention, working memory and perceived stress through completion of cognitive games. An advantage of this method is that it might remove response biases related to social acceptance. The field of crowdsourcing in music interventions has recently shown evidence of good potential but there are still several open questions on how to personalise music interventions online to enhance their acceptance and efficiency.  
\\\\
\added[id=add]{An important implication of the connection between auditory and reward systems, is that this motivates a reinforcement learning framework that influences the predictive coding model. Recently, it has been demonstrated that neuronal pathways between auditory and reward system drive learning by musically elicited `reward prediction errors' \cite{Belfi2020}. This is demonstrated in the individual differences in learning rates and explains why music-personalisation can play an important role in music-based interventions. In patients with Alzheimer disease, it was suggested that the acute arousal and consequently enjoyment in patient-selected music resulted in greater benefits in cognition and behaviour compared to clinician-chosen music \cite{Leggieri2019}.} 

However, the notion of enjoyment is complex in itself. It might naively assumed that listeners would not \added[id=add]{express musical pleasure from a piece} that evoked a negative emotional response, but as Vuust et al., 2022 \cite{Vuust} suggests, valence and enjoyment are not equated. We cannot presume that just because a piece of music evokes a negatively valence emotion, the listener does not enjoy it. In fact, sadness has been cited as the eighth most common emotion \cite{Vuust}, with reports that it provides a cathartic release therefore eases feelings of aggression \cite{Lonsdale}. This dissociation supports the important role of predictability in enjoyment rather than specific emotion evoked. Furthermore, evidence from cardiovascular physiology suggests that emotional arousal has stronger effects than emotional valence of the underlying music piece \cite{Kulinski2022}. Therefore, it might be more appropriate and effective to explore the notion of ‘enjoyment’ as opposed to a specific emotional response. 

This is an easier measurement for participants to report, since \deleted{it is a binary option and} it does not require a complex mapping between stimulus and \added[id=add]{affect}. 
Additionally, it can be measured through physiological responses that reflect increased heart rates and perspiration \cite{Vuust}. These responses are also associated with activations in the reward system such as dopamine release in the striatal system - including the caudate associated with anticipation and the NACC correlated with feelings of reward \cite{Vuust}. Secondly, it is arguably easier to manipulate the musical `grammar' or `syntax' to create desired effects of expectations violated or fulfilled while listening to music and therefore notions of enjoyment.
\\\\
In general, the field of music therapy \deleted{relatively young and therefore} arguably lacks standardisation. Different associations emphasise different \deleted{concepts in it's} requirements such as the involvement of a qualified music therapist or health practitioner to supervise sessions \cite{Steen}. Consequentially, this has led to some literature adopting the term \textbf{Music-Supported Therapy (MST)} as often studies do not comprise of a qualified music therapist.
Furthermore, \deleted{as a result of an absence of a universally-accepted defined set of musical features that can consistently evoke an emotional response,} a \added[id=add]{`researcher-friendly'} dataset of musical extracts does not currently exist for widespread use \cite{Chaturvedi}.  One step towards a more standardised proposed solution is the employment of existing datasets related to musical mood, allowing ease of comparison of validity in results and methodology \cite{Chaturvedi}.

\section{Conclusions}
In the field of music therapy, the use of music has been shown to exceed it's archetypal characteristic of simply mood regulation and entertainment purposes. Advances in use of music to support neuro- and physical rehabilitation proves that it is a compelling contender to current interventions used in the healthcare sector. Building from models of rhythm perception and processing, concepts of predictive-coding and oscillatory relationships between neural networks has been adopted to entrain patterns of stability in gait abnormalities, promoting faster and less variable strides. Further positive secondary outcomes include improvement in overall quality of life, feelings of motivation and reduction in feelings of stress. Despite promising results, it was also discussed the importance tailoring interventions, including the initial pulse of rhythmic stimuli to match participants natural cadence, such that maximal gain can be extracted. Choosing the `correct' musical stimuli was shown to be of crucial to avoid increase in cognitive load and causing mental fatigue. However, this was noted to have it's own additional challenges. The development of effective interventions that do not rely on medication and can be transferred into the community easily provides an encouraging prospect into long-term treatment that can eventually replace or complement traditional methods.

\section{ACKNOWLEDGMENT}

The authors acknowledge funding by an EPSRC new investigator award (EP/W01212X/1), Royal Society (RGS/R2/212199) and the UKRI Centre for Doctoral Training in Socially Intelligent Artificial Agents (EP/S02266X/1)

\bibliographystyle{ieeetr}
\bibliography{References_IEEE.bib}

\end{document}